\documentclass[
reprint,
superscriptaddress,
 amsmath,amssymb,
 aps,
pra,
]{revtex4-2}

\usepackage{graphicx}
\usepackage{dcolumn}
\usepackage{bm}
\usepackage{wasysym}
\usepackage{xcolor}
\usepackage{float}
\newcommand{\be} {\begin{equation}}
\newcommand{\ee} {\end{equation}}
\newcommand{\ben} {\begin{equation*}}
\newcommand{\een} {\end{equation*}}
\newcommand{\eq}[1]{\begin{align}#1\end{align}}
\newcommand{\eqn}[1]{\begin{align*}#1\end{align*}}

\newcommand{\ba} {\begin{array}}
\newcommand{\ea} {\end{array}}
\newcommand{\citeasnoun}[1]{Ref.~\citenum{#1}}
\renewcommand{\vec}[1]{\boldsymbol{#1}}

\newcommand{\Figref}[1]{Figure~\ref{#1}}

\begin{document}

\title{Reduced model for capillary breakup with thermal gradients: \\ Predictions and computational validation}

\author{Isha Shukla}
\affiliation{Laboratory of Fluid Mechanics and Instabilities, Ecole Polytechnique F\'ed\'erale de Lausanne, 1015 Lausanne, Switzerland}
\affiliation{Department of Mechanical and Aerospace Engineering, University of California, San Diego, La Jolla, CA 92093, USA}

\author{Fan Wang}
\affiliation{Department of Mathematics, Massachusetts Institute of Technology, Cambridge, MA 02139, USA}

\author{Saviz Mowlavi}
\affiliation{Department of Mechanical Engineering, Massachusetts Institute of Technology, Cambridge, MA 02139, USA}

\author{Amy Guyomard}
\affiliation{Department of Mathematics, Massachusetts Institute of Technology, Cambridge, MA 02139, USA}

\author{Xiangdong Liang}
\affiliation{Department of Mathematics, Massachusetts Institute of Technology, Cambridge, MA 02139, USA}

\author{Steven G. Johnson}%
\affiliation{Department of Mathematics, Massachusetts Institute of Technology, Cambridge, MA 02139, USA}

\author{J.-C. Nave}
\affiliation{Department of Mathematics, McGill University, Montr\'eal, Qu\'ebec H3A 0B9, Canada}


\date{\today}
             
\begin{abstract}
It was recently demonstrated that feeding a silicon-in-silica coaxial fibre into a flame---imparting a steep silica viscosity gradient---results in the formation of silicon spheres whose size is controlled by the feed speed [Gumennik et al., Nat.~Commun.~4, 2216 (2013)]. A reduced model to predict the droplet size from the feed speed was then derived by Mowlavi et al.~[Phys.~Rev.~Fluids.~4, 064003 (2019)], but large experimental uncertainties in the parameter values and temperature profile made quantitative validation of the model impossible. Here, we validate the reduced model  against fully-resolved three-dimensional axisymmetric Stokes simulations using the exact same physical parameters and temperature profile. We obtain excellent quantitative agreement for a wide range of experimentally relevant feed speeds. Surprisingly, we also observe that the local capillary number at the breakup location remains almost constant across all feed speeds. Owing to its low computational cost, the reduced model is therefore a useful tool for designing future experiments.
\end{abstract}

\maketitle
\section{Introduction}

\begin{figure}
    \centering
    \includegraphics[width=0.5\textwidth]{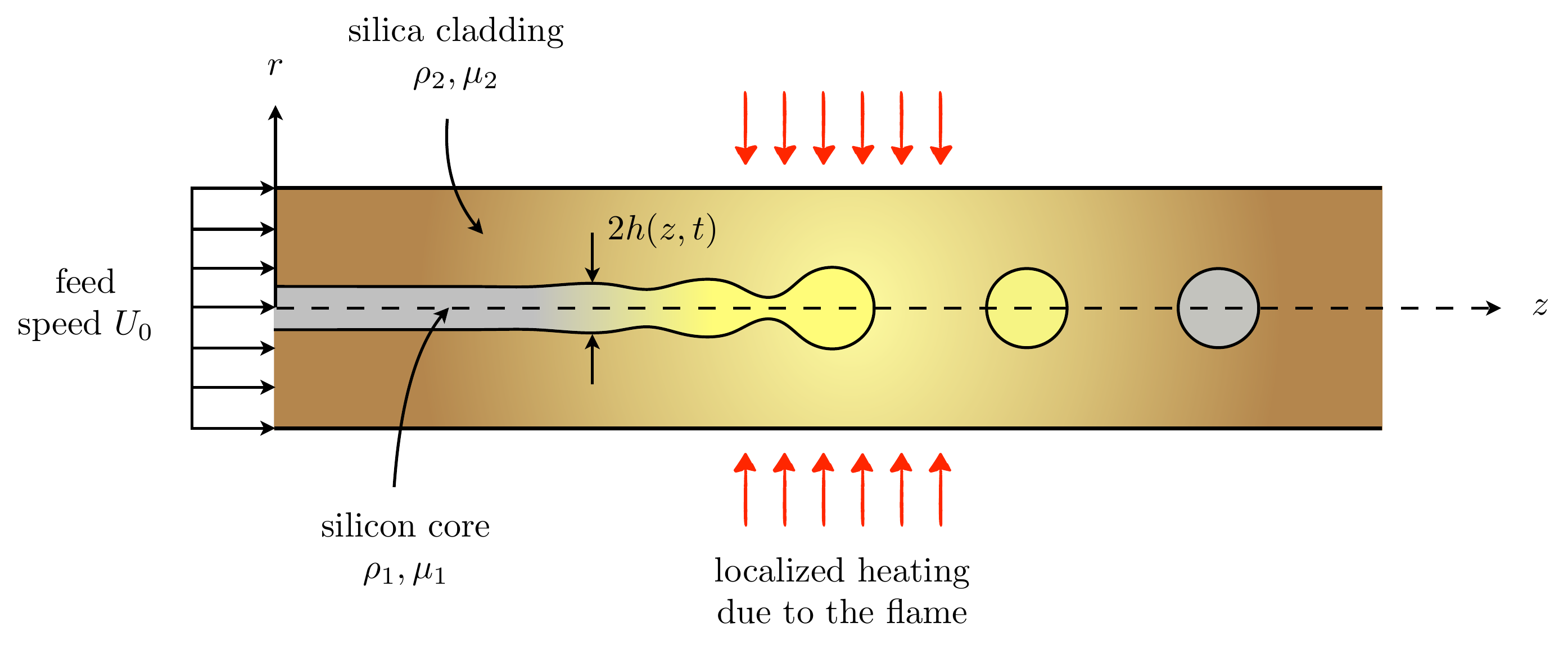}
    \caption{Schematics of the problem. A silicon-in-silica co-axial fiber is fed through a flame at a constant speed, imparting an axial thermal gradient along the fibre. The ensuing melting of the silicon core (pictured by the transition from gray to yellow color) together with the softening of the silica cladding (pictured by the shift from darker to lighter brown color) triggers a capillary breakup of the silicon core into regular spheres. Note that the colors do not reflect the actual values of the viscosity}
    \label{fig:fig1}
\end{figure}

The classic phenomenon of capillary breakup of a jet into droplets~\cite{eggers2008physics} has recently been revisited under a new experimental setting: a fibre is fed through a steep thermal/viscosity gradient (\Figref{fig:fig1}), where the feed speed provides control over the droplet size~\cite{gumennik2013silicon}.  In that initial work, a silicon-in-silica co-axial fibre was fed into a localized flame, breaking up into spherical silicon droplets far smaller~\cite{gumennik2013silicon} than those produced by a classic isothermal process~\cite{KaufmanTa12}.  However, a key challenge has been to develop a simplified model that quantitatively predicts the droplet size from the feed speed and other parameters, in order to better understand this phenomenon and to design future experiments.

In~\citeasnoun{mowlavi2019particle}, such a one-dimensional (1D) reduced model was developed through a long-wavelength approximation of the governing Navier--Stokes equations. However,  large experimental uncertainties in the temperature/viscosity profile made it impossible to quantitatively validate the model's accuracy. In the present work, we circumvent this difficulty by validating the 1D reduced model against fully-resolved three-dimensional (3D) axisymmetric Stokes simulations of the same problem. We observe excellent quantitative agreement for a wide range of experimentally relevant feed speeds ($U_0=5$--$30\,\mu\mathrm{m}/s$). 
This is consistent with our observation that the local capillary number at the breakup location remains constant irrespective of the feed speed. Because the capillary number scales as the product of feed speed and viscosity, sufficiently high feed speeds cause breakup to occur deeper into the flame, in such a way that the long-wavelength assumption of the reduced model holds. As a result, the reduced model is a useful tool for designing future experiments, since it takes the form of a pair of 1D partial differential equations (PDEs)~\cite{mowlavi2019particle} that are vastly simpler and more efficient to simulate than a full 3D axisymmetric Stokes model.


Capillary instability of liquid threads and jets is a widely studied subject. It was Plateau~\cite{plateau1873statique} who, based on geometric arguments, first demonstrated that a liquid cylindrical thread would become unstable due to capillary forces when the thread length exceeds its circumference. Later, Lord Rayleigh~\cite{rayleigh1879capillary,rayleigh1892xvi} used linear-stability analysis to obtain the value of the most unstable breakup mode for an inviscid jet, leading to a quantitative prediction of the resulting drop size. Rayleigh's analysis was later generalized by Tomotika to a coaxial cylinder consisting of two fluids with different viscosities~\cite{tomotika1935instability}, and was more recently extended further to an arbitrary number of concentric fluids with different viscosities and densities~\cite{liang2011linear}. 
%
These linear-stability analysis tools give accurate predictions for the size of particles produced by isothermal co-axial fibre-drawing processes~\cite{KaufmanTa12}, but they do not predict the droplets in the dynamic thermal-gradient process considered here~\cite{gumennik2013silicon,mowlavi2019particle}. Simple extensions of linear-stability analysis to the thermal-gradient case (\Figref{fig:fig1}) have thus far failed to produce accurate results~\cite{mowlavi2019particle} or require unknown dimensionless fit parameters~\cite{gumennik2013silicon}. This motivated the numerical simulations of the long-wavelength reduced model presented in~\citeasnoun{mowlavi2019particle}, which were in reasonable agreement with experimental results from~\citeasnoun{gumennik2013silicon}. Nonetheless, large uncertainties in the experimental temperature profile, which lead to exponentially large uncertainties in the viscosity~\cite{sato2003viscosity,doremus2002viscosity}, made it impossible to precisely validate the reduced model. Reducing these uncertainties in future experiments will be an arduous process. The alternative is to validate against brute-force Stokes simulations (valid because the relevant Reynolds number is on the order of $10^{-13}$), which offer a precise comparison with exactly known parameters.

We close this introduction by giving a brief description of the two models that we compare in this paper, both applied to the problem pictured in \Figref{fig:fig1}. The reduced model from~\citeasnoun{mowlavi2019particle}, which is obtained from a long-wavelength approximation of the Euler equations for the silicon core and the Stokes equations for the silica cladding, takes the form of two coupled 1D PDEs that we solve in Matlab. 
Separately, a large-scale parallel solver for the 3D axisymmetric Stokes equations was developed in~C, giving an accurate reference solution for the capillary breakup process and resulting particle size. For various temperature profiles and feed speeds, we obtain different particle sizes, which are then compared with predictions from the 1D reduced model to demonstrate the accuracy and determine the range of applicability of this model.  Because solving the reduced model is much faster than a full Stokes simulation---even with our unoptimized Matlab code---the validated reduced model is therefore a useful tool for designing experiments and extracting future analytical insights.

\section{Modeling}

\subsection{Problem setup}

The problem that we consider throughout this paper, inspired from the experimental setup of Gumennik et al.~\cite{gumennik2013silicon}, is pictured in \Figref{fig:fig1}. A coaxial fiber made of a silicon core of radius $h_0 = 2 \, \mu\mathrm{m}$ encased in a much larger silica cladding is fed into a localized flame at a uniform speed $U_0$. The local temperature gradient imparted by the flame causes the silicon core to melt and the silica cladding to soften, triggering capillary breakup of the silicon core into a continuous string of spheres. These silicon spheres solidify upon leaving the flame and remain trapped within the silica matrix.

This study concerns the region downstream of the liquefaction point of the silicon core, which we set as the origin of the axial coordinate $z$. The fiber witnesses temperatures ranging from $T_l \simeq 1400^{\circ}$C at the melting point of silicon to $T_h \simeq 1850^{\circ}$C in the heart of the flame~\cite{gumennik2013silicon}, but experimental limitations prevented the measurement of a detailed temperature profile. For validation purposes, therefore, a hyperbolic tangent profile is assumed:
\eq{\label{Temp_eqn}
T(z) = T_l + (T_h - T_l) \tanh \left(\frac{z}{w} \right),
}
where $w$ is the length scale associated with the temperature gradient. Over this temperature range, the molten silicon core and silica cladding have relatively constant density $\rho_\mathrm{i} \simeq \rho_\mathrm{o} \simeq 2500 \, \mathrm{kg/m}^3$. However, their viscosities depend exponentially on the temperature according to the following functions~\cite{doremus2002viscosity,sato2003viscosity}:
\eq{
\label{mu_Si}
\mu_\mathrm{i}(z) &= 10^{\frac{819}{T(z)+273}-3.727} \, \mathrm{Pa \, s}, \\
\label{mu_SiO2}
\mu_\mathrm{o}(z) &= 10^{\frac{26909}{T+273}-7.2348} \, \mathrm{Pa \, s}.
}
As a result, the viscosity profile $\mu_\mathrm{i}(z)$ of the inner silicon varies from $6 \cdot 10^{-4}$ to $5 \cdot 10^{-4}  \, \mathrm{Pa \, s}$, while the viscosity profile $\mu_\mathrm{o}(z)$ of the outer silica varies from $7 \cdot 10^8$ to $3 \cdot 10^5  \, \mathrm{Pa \, s}$. Although the viscosity of the inner silicon is not significantly affected by the axial thermal gradient imposed by the flame, the opposite is true for the outer silica, whose viscosity changes by more than three orders of magnitude over millimeter scales. Such a drastic viscosity gradient is believed to be the reason for the dependence on feed speed of the resulting droplet size~\cite{gumennik2013silicon,mowlavi2019particle}. Finally, the surface tension between silicon and silica is taken to be $\gamma = 10 \, \mathrm{N/m}$.

Before presenting the two models that are considered in this study, we summarize the physical mechanisms dominating the dynamics of this problem. Besides surface tension~$\gamma$, which clearly plays an important role, there are four material parameters: the densities $\rho_\mathrm{i}$ and $\rho_\mathrm{o}$ of the silicon and silica, and their viscosities $\mu_\mathrm{i}(z)$ and $\mu_\mathrm{o}(z)$. In the dimensional analysis to follow, we will consider the viscosity values corresponding to the highest temperature point in the heart of the flame. The Reynolds number in the outer silica, $\mathrm{Re}_\mathrm{o} = \rho_\mathrm{o} U_0 h_0/\mu_\mathrm{o} \sim 10^{-13}$, shows that inertial effects due to $\rho_\mathrm{o}$ are negligible. The dynamics of the inner silicon are governed by surface tension rather than by the velocity scale associated with the feed speed of the fiber. Thus, the relevant quantity to compare the relative importance of viscous and inertial effects in the silicon is the Ohnesorge number, defined as the ratio of viscous to inertial timescales of the capillary instability~\cite{mckinley2011wolfgang}. We find that $\mathrm{Oh}_\mathrm{i} = \mu_\mathrm{i}/\sqrt{\rho_\mathrm{i} \gamma h_0} \sim 10^{-3}$, revealing that viscous effects in the silicon are negligible. We are then left with $\rho_\mathrm{i}$ and $\mu_\mathrm{o}$, which we compare using a mixed Ohnesorge number $\mathrm{Oh}_{\mathrm{i}/\mathrm{o}} = \mu_\mathrm{o}/\sqrt{\rho_\mathrm{i} \gamma h_0} \sim 10^6$. The latter indicates that, ultimately, viscosity of the outer silica $\mu_\mathrm{o}$ is the only meaningful material parameter acting together with surface tension.

Consequently, both models considered in this study take into account the spatially-varying silica viscosity $\mu_\mathrm{o}(z)$. Although the inertia and viscosity of the inner silicon are both negligible, it turns out that it is necessary to include at least one of them in order to have well-posed governing equations. The two models make different choices in this regard---the axisymmetric Stokes solver clearly neglects inertia of the inner silicon, while the reduced model is derived from the Euler equations for the inner silicon, which do not account for viscosity. Nevertheless, both approaches are physically meaningful since the equations are then simulated in a range of parameter values for which the only mechanisms that matter are the surface tension and viscous dissipation in the outer silica.

\subsection{Full Stokes model}
We begin with the description of the full 3D axisymmetric Stokes solver. The dynamics are described by the Stokes equation
\be
-\nabla p + \nabla \cdot \left[ \mu(\vec{r})\left( \nabla \bm{v} + \nabla \bm{v}^T \right) \right] =  
\gamma\, \delta( \phi(\vec{r}) ) \, \kappa( \phi(\vec{r}) ) \, \frac{\nabla  \phi(\vec{r}) }{\left|\nabla  \phi(\vec{r} ) \right|}
\label{stokes}
\ee
and the continuity equation
\be
\nabla \cdot \bm{v} = 0,
\label{eq:continuity}
\ee
where $p$ and $\bm{v}$ are the pressure and velocity fields in both the inner and outer layers, $\phi(\vec{r})$ is the level-set function which defines the interface position $\vec{r}_\mathrm{i}$ through $\phi(\vec{r}_\mathrm{i})=0$, and $\kappa(\phi(\vec{r}_\mathrm{i}))$ is the curvature of the interface. The motion of the interface is described by the advection equation
\be
\frac{\partial \phi}{\partial t} + \bm{v} \cdot \nabla \phi = 0
\label{tevol}
\ee

Numerically, \eqref{stokes} and \eqref{eq:continuity} are discretized by a second-order finite-difference scheme, and solved by a parallel MUMPs direct solver~\cite{amestoy2000multifrontal} using the PETSc library~\cite{petsc-user-ref}. For \eqref{tevol} we use a third-order TVD Runge-Kutta method for time integration and the WENO discretization in space~\cite{sussman1994level}. More details on the numerical schemes are presented in the Appendix. The simulation domain extends 30\,$\mu$m in the radial direction with grid spacing of 0.2\,$\mu$m, and the length in the axial direction is $2 \, \mathrm{mm}$.

%
\subsection{Reduced model}
The reduced model, introduced in~\citeasnoun{mowlavi2019particle}, is derived from a long-wavelength approximation of the incompressible Euler equations for the inner silicon and the incompressible Navier--Stokes equations for the outer silica. It consists of a set of two coupled 1D nonlinear PDEs for the leading-order inner velocity $u_\mathrm{i}(z)$ and interface height $h(z)$. Expressed in dimensionless form using the length scale $h_0$ and velocity scale $U_0$, the coupled 1D equations take the form
\begin{subequations}\label{eqN2}
\begin{align}
\mathrm{We} \left(\frac{\partial v}{\partial \tilde{t}} + v\frac{\partial v}{\partial \tilde{z}}\right ) &=  - \frac{\partial \tilde{\kappa}}{\partial \tilde{z}} - \frac{\partial}{\partial \tilde{z}} \left[ \frac{\mathrm{Ca}_{\tilde{z}}}{f} \left(-\frac{\partial (fv)}{\partial \tilde{z}} + \frac{\partial f}{\partial \tilde{z}}\right) \right],
\\
\frac{\partial f}{\partial \tilde{t}} &= -\frac{\partial (fv)}{\partial \tilde{z}},
\\ 
\tilde{\kappa} &= \frac{(2-f'')f + f'^2}{2 ( f'^2/4 + f)^{3/2}},
\end{align}
\end{subequations}
where $\tilde{z} = z/h_0$, $\tilde{t} = tU_0/h_0$, $v = u_\mathrm{i}/U_0$ is the dimensionless velocity, $f = (h/h_0)^2$ is a dimensionless function describing the interface radius, $\tilde{\kappa}$ is the dimensionless interface curvature, and $\mathrm{We} = {\rho_\mathrm{i} h_0 {U_0}^2}/{\gamma}$ and $\mathrm{Ca}_{\tilde{z}} = {\mu_\mathrm{o}(\tilde{z}) U_0}/{\gamma}$ are respectively the Weber and spatially-varying capillary numbers.

The Weber number $\mathrm{We}$ based on the true values for the physical parameters lies in a range that is computationally inaccessible. Nonetheless, it was demonstrated in~\citeasnoun{mowlavi2019particle} that below a certain limit, the Weber number has a negligible influence on droplet size. We henceforth pick $\mathrm{We} = 0.05$ in our simulations, regardless of the feed speed $U_0$. On the other hand, the capillary number $\mathrm{Ca}_{\tilde{z}}$ is calculated from equations \eqref{mu_SiO2} and \eqref{Temp_eqn}, and therefore inherits a large spatial gradient from the silica viscosity $\mu_\mathrm{o}(z)$.

A numerical domain of length 3\,mm with a grid spacing of 0.5--1\,$\mu$m is considered. For smaller feed speeds $U_0$ ranging from 1 to 5\,$\mu$m/s, the initial condition is defined as a cylinder extending 0.22\,mm into the domain and with a unit dimensionless radius. The cylinder is enclosed with a spherical tip and its dimensionless interface velocity is set to~1. Beyond the cylinder tip, the height and velocity are set as~0. For feed speeds larger than $U_0=5\,\mu$m/s, simulations are performed sequentially at intervals of $5\,\mu$m/s, beginning from $U_0=10\,\mu$m/s and terminating at $U_0=30\,\mu$m/s. As individual simulations are run for these increasing feed speeds, the convergence time is improved by selecting as initial condition the breakup shape and velocity obtained from the preceding feed speed.
We verified that the initial shape of the jet does not affect its quasi-steady breakup characteristics, which are the focal point of our analysis. A maximum breakup length of approximately 1.3\,mm is observed, which is within the limits of the domain size. The simulations are run for a sufficiently long time to enter the regime where the jet breaks up at regular intervals of time and at a fixed axial location. The jet characteristics, such as the breakup length, location and drop radius, are obtained in this quasi-steady regime. When satellite drops appear, they are equal to about $3.3\%$ (for $U_0>5\mu$m$/$s) and $10\%$ (for $U_0<5\mu$m$/$s) of the main drop volume and hence are considered as numerical artifacts.
%
\begin{figure*}
    \centering
    \includegraphics[width=0.55\textwidth]{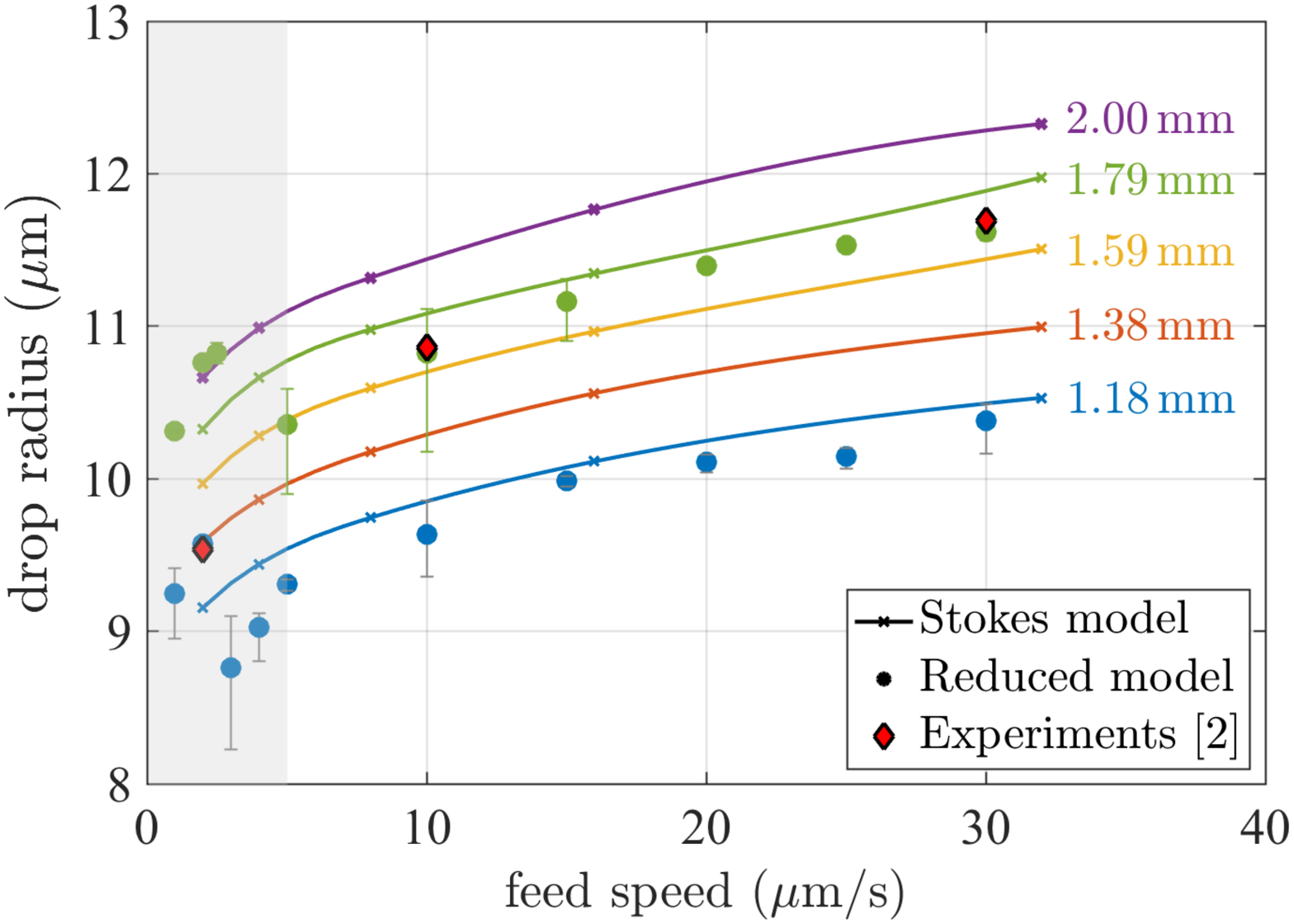}
    \caption{Comparison of drop radii obtained by the full Stokes model (lines), the semi-analytical 1D model (circles), and  experimental measurements~\cite{gumennik2013silicon} (diamonds) as a function of the feed speed. The solid line for the full Stokes model is obtained using a spline fit of the actual data  obtained for feed speeds $U_0=2,4,8,16~\text{and}~32$\,$\mu$m$/$s. Different transition widths $w$ ranging from 1.18\,mm to 2.0\,mm are considered for the temperature profile of the hydrogen torch, which varies from 1400$^{\circ}$C to 1850$^{\circ}$C. The experimental measurements correspond to the numerical results for $w \simeq 1.16$\,mm.  In the shaded (low-speed) region, the long-wavelength approximation breaks down and the 1D reduced model is no longer accurate.}
    \label{fig:fig3}
\end{figure*}
\begin{figure*}
    \centering
    \includegraphics[width=0.7\textwidth]{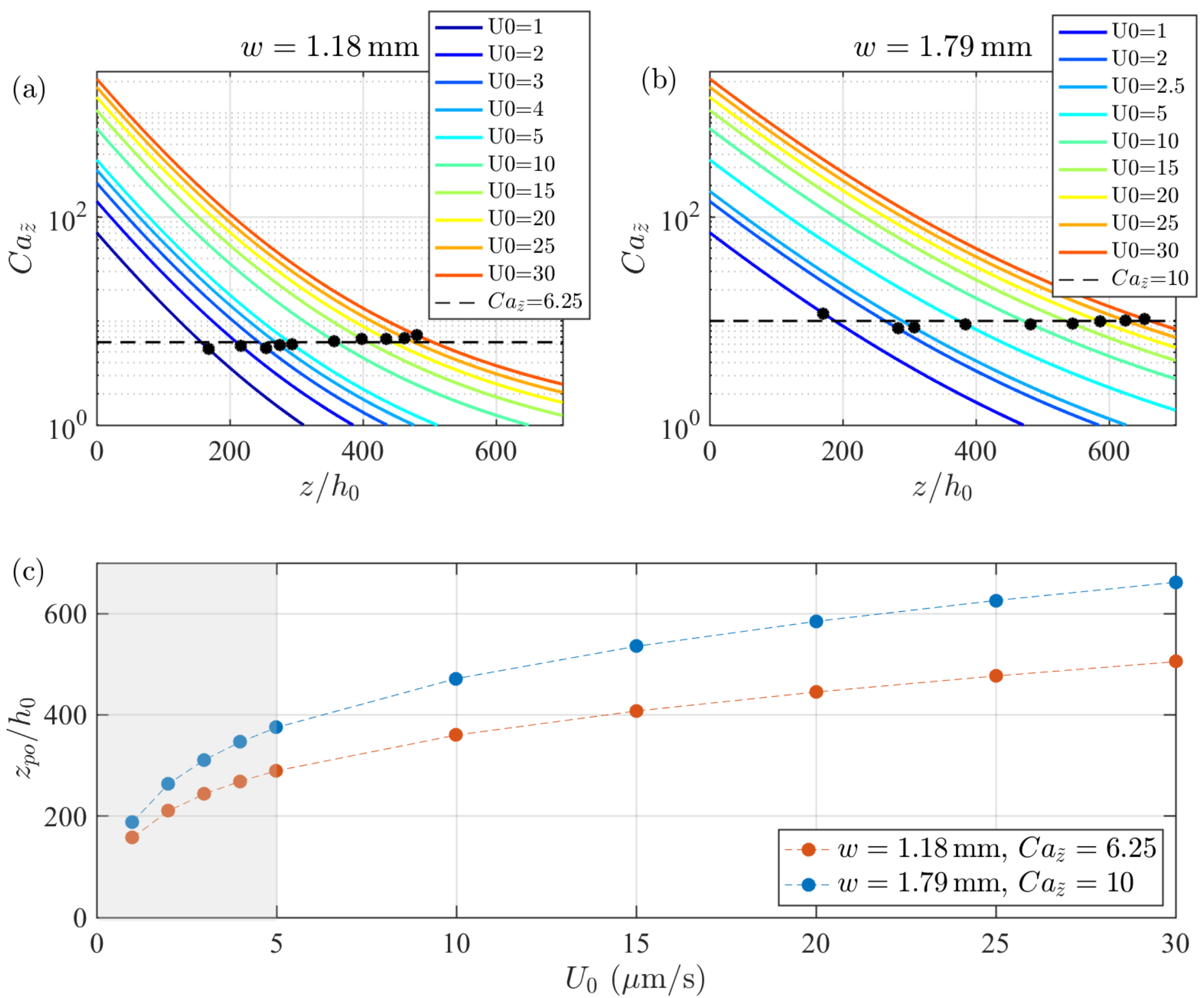}
    \caption{Capillary number profile as a function of the dimensionless axial distance for (a) $w=1.18$\,mm and (b) $w=1.79$\,mm, and for feed speeds between 1--30\,$\mu$m/s. The black markers represent the axial location where the silicon jet breaks up into drops and the corresponding capillary number at that location. (c) Dimensionless breakup location inferred from the mean capillary number at breakup, as a function of the feed speed and for different widths. Shaded region indicates where the long-wavelength assumption of the reduced model breaks down due to a rapid decrease of the breakup location at low speeds.}
    \label{fig:fig4}
\end{figure*}
\section{Results and discussion} 
The sphere radius as a function of the feed speed, obtained using both the full Stokes model and the reduced order 1D model, are plotted in Fig.~\ref{fig:fig3} . The results in bold lines are obtained from the 3D Stokes solver with different values of temperature transition width, $w$, ranging from 1.2\,mm to 2.0\,mm; and in colored circular markers, obtained by numerically solving the 1D equation \eqref{eqN2} for $w=1.18$ and 1.79\,mm only. Good quantitative agreement for the drop radius is seen between the two approaches in the speed range $U_0=5$--$30\,\mu\mathrm{m}/\mathrm{s}$. Additionally, we also plot as red diamond markers the radii obtained from the experimental measurements~\cite{gumennik2013silicon}. The 3D Stokes results agree quite well with the experimental results for feed speeds as low as $U_0=10\,\mu\mathrm{m}/\mathrm{s}$ and suggest an experimental value of $w$ around $1.6 \, \mathrm{mm}$, which is within the range of expected experimental values for a hydrogen torch.

Fig.~\ref{fig:fig3} shows that the sphere radius has a nonlinear dependence on the feed speed. As the feed speed increases, the mass flow rate of silicon into the high temperature region increases, thereby adding more volume to the sphere before breakup. However the increase in feed speed also shifts the pinch-off location further down into the higher temperature region, thereby decreasing the pinch-off time and reducing the amount of silicon that is used in the formation of the sphere. 

Below $U_0=5\,\mu$m/s (see the grey shaded region in Fig. \ref{fig:fig3}) a deviation is observed between the drop radii obtained from the 1D solver and from the Stokes solver.
In the shaded region, the large variation in the drop radius predicted by the 1D model can be probably attributed to the limitations of the numerical scheme and the failure of the long-wavelength assumption used in deriving the simplified governing equations \eqref{eqN2}. The latter reason is further investigated by plotting the pinch-off location for different feed speeds. First, we plot the spatial profiles of the capillary number $\mathrm{Ca}_{\tilde{z}}$ for the different feed speeds as a function of the dimensionless axial coordinate $z/h_0$ in Figs.~\ref{fig:fig4}(a) and \ref{fig:fig4}(b) for $w=1.18$\,mm and 1.79\,mm, respectively. In the same plot we also indicate, in black circular markers, the pinch-off location observed for each feed speed. Surprisingly, for a given value of width $w$, the capillary number at the pinch-off location, $\mathrm{Ca}_{\tilde{z},po}$, was found to be almost constant irrespective of the feed speed. We find the mean values $\langle \mathrm{Ca}_{\tilde{z},po}\rangle = 6.25$ for $w=1.18$\,mm and $\langle \mathrm{Ca}_{\tilde{z},po}\rangle = 10$ for $w=1.79$\,mm, which we indicate as the dashed lines in Fig.~\ref{fig:fig4}(a) and \ref{fig:fig4}(b). Using these mean values, we may infer the breakup location $\tilde{z}_{po}$ through the definition of the capillary number as
\be
z_{po} = h_0 \mu_\mathrm{o}^{-1} \left( \frac{\gamma \langle \mathrm{Ca}_{\tilde{z},po} \rangle}{U_0} \right),
\ee
where $\mu_\mathrm{o}^{-1}$ is the inverse of equation \eqref{mu_SiO2} for the silica viscosity profile. The resulting break-up locations $\tilde{z}_{po} = z_{po}/h_0$ are plotted in Fig.~\ref{fig:fig4}(c) with respect to the feed speed. Interestingly, the combination of an exponentially decreasing viscosity profile together with a constant capillary number at breakup leads $z_{po}/h_0$ to decrease monotonically as $U_0$ is reduced. This decrease, however, is steepest below a feed speed of $5 \, \mu$m/s. For example, $z/h_0$ decreases by an order of $150 \, \mu$m when $U_0$ reduces from 30 to $10 \, \mu$m/s and by a similar order for $U_0$ from 5 to $1 \, \mu$m/s. Thus, the long-wavelength assumption used by the 1D model could be violated for feed speeds $U_0 \lesssim 5 \, \mu$m/s where the breakup length is smallest, possibly explaining the failure of the 1D model in that regime.

\section{Concluding remarks}
In this paper, we validated a 1D reduced model that predicts the sphere size formed by capillary breakup in the presence of steep temperature and viscosity gradients. The reduced model was introduced in~\citeasnoun{mowlavi2019particle}, but large uncertainty in the experimental temperature profile made its precise validation impossible. This issue is addressed here by validating the model against fully-resolved 3D axisymmetric Stokes simulations of the same problem. Without any adjustable parameters, the 1D model accurately estimates the sphere size in the experimentally relevant range of feed speeds ($U_0=5$--$30\,\mu$m/s). For a fixed lengthscale $w$ of the temperature gradient, we observe near-constancy of the capillary number at the breakup location, irrespective of the feed speed, which is an experimentally relevant piece of information in gauging the breakup lengths for arbitrary feed speeds. Owing to its significantly lower computational cost, the 1D reduced model provides a useful tool for designing experiments and for gaining future physical insights. 

However, for the given temperature profile, the 1D reduced model is unreliable for feed speeds below $U_0=5\,\mu$m/s, where the long-wavelength assumption of the model is expected to be violated. Thus, predictions of droplet radii in this range of feed speeds should be obtained using the full 3D Stokes solver, unless and until the reduced model can be refined so as to work in this regime. Furthermore, the numerical implementation of the reduced model can benefit from multiple improvements. Our current numerical scheme handles the splitting of the jet and the motion of the tip by checking for conditions that depend on different numerical parameters which need to be adjusted manually. Using a different scheme devoid of such parameters, for instance by borrowing ideas from the regularized approach introduced in~\citeasnoun{driessen2011regularised}, would make it easier to run the model over a wider range of physical settings. Finally, interesting future work could include adapting the numerical scheme to capture the evolution of the drop beyond its breakup, as it enters a region of decreasing temperature. This would be extremely relevant for experimental conditions where the temperature profile solidifies the droplet soon after  breakup, preventing its complete transformation into a sphere.
\begin{acknowledgments}
IS and SM wish to thank Fran\c cois Gallaire and P.-T.~Brun for useful discussions during the development of the reduced 1D model. JCN acknowledges financial support through the NSERC Discovery Grant program. FW and SGJ were supported in part by the MRSEC Program of the National Science Foundation under award number DMR-1419807.
\end{acknowledgments}
\appendix

\section{Stokes discretization}
The governing equations for capillary breakup in a concentric two-phase fluid system are \eqref{stokes}--\eqref{tevol} in the main text.
While advecting the level-set function $\phi$, contours in the vicinity of the $\phi=0$ may become increasingly distorted, thus leading to potentially large error in evaluating derivatives. In order to mitigate this issue, a reinitiallization process aiming to restore the $|\nabla \phi|=1$ property in the neighborhood of zero level set is necessary. This is achieved by solving in pseudo-time $\tau$ the reinitialization equation,
\eq{
\label{app_renorm}
\frac{\partial \phi}{\partial \tau} + \mbox{sign}(\phi_0) (|\nabla \phi| -1) = 0,
}
where $\mbox{sign}(\phi_0)$ is a smoothed sign function evaluated from the level-set function at $\tau=0$. For details on the approach and the discretization we use, we refer the reader to~\citeasnoun{sussman1994level}.

Since the capillary breakup problem possesses azimuthal symmetry, we express the governing equations in cylindrical coordinate ($r,z$). In particular, the explicit form of the left-hand side of \eqref{stokes}, assuming $\bm{v} = u\hat{\bm{e}}_r + w \hat{\bm{e}}_z$, is
\eqn{
 &-\nabla p + \nabla \cdot \left[ \mu \left( \nabla \bm{v} + \nabla \bm{v}^T \right) \right] =\\
 &\left\{ -\frac{\partial p}{\partial r} + \frac{2}{r} \frac{\partial}{\partial r} \left( \mu r\frac{\partial u}{\partial r} \right) {{- 2\mu \frac{u}{r^2 }} }+ \frac{\partial }{\partial z} \left[ \mu \left( \frac{\partial w}{\partial r} + \frac{\partial u}{\partial z} \right) \right]  \right\} \hat{\bm{e}}_r \\
&+\left\{ -\frac{\partial p}{\partial z} + \frac{1}{r} \frac{\partial }{\partial r} \left[ \mu r \left( \frac{\partial u}{\partial z} + \frac{\partial w}{\partial r} \right) \right] + \frac{\partial}{\partial z} \left( 2 \mu \frac{\partial w}{\partial z} \right) \right\} \hat{\bm{e}}_z.
\label{lhsstokes}
}
The continuity equation \eqref{eq:continuity} becomes
\be
\nabla \cdot \bm{v} = \frac{1}{r} \frac{\partial}{\partial r} (ru) + \frac{\partial w}{\partial z} = 0.
\ee
From the level-set function, we compute the unit normal vector 
\eq{
\bm{n} = \frac{\nabla \phi}{|\nabla \phi|} = \frac{\phi_r}{\sqrt{\phi_r^2 + \phi_z^2}} \hat{\bm{e}}_r + \frac{\phi_z}{\sqrt{\phi_r^2 + \phi_z^2 }} \hat{\bm{e}}_z,
}
where $\phi_r=\partial \phi/\partial r$ and $\phi_z = \partial \phi /\partial z$. Additionally, the curvature is defined as $\kappa = \nabla \cdot \bm{n}$, hence 
\eq{
&\kappa(\phi) & = \frac{\phi_{rr}\phi_z^2 - 2\phi_r\phi_z\phi_{rz} + \phi_{zz}\phi_r^2}{\left( \phi_r^2 + \phi_z^2 \right)^{3/2} } + \frac{1}{r} \frac{\phi_r}{\sqrt{\phi_r^2 + \phi_z^2}}.
}
Furthermore, we require a smoothed Dirac delta function $\delta_\epsilon$, which we define as 
\be
\delta_\epsilon(\phi)= \left\{ 
\begin{array}{ll}
0, & \quad \mbox{for } \phi \le -\epsilon,\\
\left[1+\cos\left( \pi \phi / \epsilon\right)\right] / 2\epsilon, & \quad \mbox{for } -\epsilon < \phi < \epsilon,\\
0, & \quad \mbox{for } \phi \ge \epsilon.
\end{array}
\right.
\label{delta_epsilon}
\ee
Note that the approximation \eqref{delta_epsilon} possesses the following desired property
\be
\delta (\phi) = \lim_{\epsilon\to0^+}\delta_\epsilon(\phi). 
\ee
In our implementation, $\epsilon = 3\Delta r$ where $\Delta r$ is the grid spacing in the $r$ direction.

\begin{figure}[ht]
\centering
\includegraphics[width=0.15\textwidth]{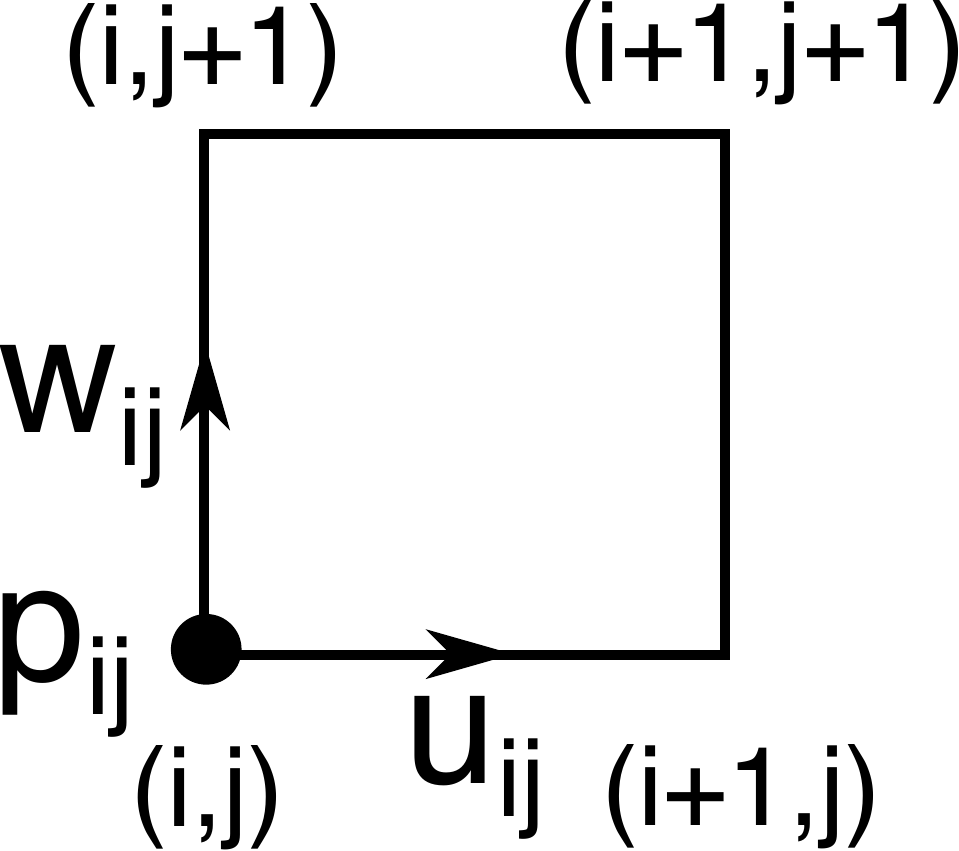}
\caption{Discretization grid of pressure $p$, velocities in $r$ direction $u$, and in $z$ direction $w$.}
\label{grid}
\end{figure}

The spatial discretization is performed on a staggered grid where the pressure is on the cell corners, the $r$-direction velocity $u$ is located on the horizontal cell interfaces, and the $z$-direction $w$ is located on the vertical cell interfaces, as shown in Fig.~\ref{grid}. The Stokes equation is discretized at grid locations $(i+1/2,j)$ for the $\hat{\bm{e}}_r$ component and $(i,j+1/2)$ for the $\hat{\bm{e}}_z$ component. The continuity equation is discretized at grid locations $(i,j)$.

Special care must taken when discretizing the $\hat{\bm{e}}_z$ component and the continuity equation at $i=0$ and $r=0$. This can be resolved by expansion around $r=0$. Rewrite
$
q(r,z) \equiv \mu \partial u / \partial z.
$
Since there is no extra source at $r=0$, we have $u(0,z)=0$ and $q(0,z)=0$, and we can express $q(r,z)$ by Taylor expansion
\eq{
q(r, z) = r \frac{\partial q(r,z)}{\partial r} + r^2\frac{\partial^2 q(r,z)}{\partial r^2} + O(r^3),
}
so that
\eqn{
\left. \frac{1}{r}\frac{\partial }{\partial r} \left( \mu r \frac{\partial u}{\partial r} \right) \right|_{r\to0} 
= \left. \frac{1}{r} \frac{\partial }{\partial r} (rq(r,z))\right|_{r\to0} 
= 2\frac{\partial }{\partial r}\left(\mu \frac{\partial u}{\partial z} \right).
}
Similarly, we obtain
\eqn{
\left. \frac{1}{r} \frac{\partial }{\partial r} \left( \mu r \frac{\partial w}{\partial r} \right) \right|_{r\to0} = 2\frac{\partial }{\partial r} \left( \mu \frac{\partial w}{\partial r} \right),
}
since $\partial w /\partial r |_{r\to0} \to 0$, and
\ben
\left.\frac{1}{r}\frac{\partial }{\partial r} (ru)\right|_{r\to0} = 2\frac{\partial u}{\partial r}.
\een

For the time evolution in \eqref{tevol} and \eqref{app_renorm}, we use a third order TVD Runge-Kutta method and the HJ WENO method for spatial discretization, following the approach in \citeasnoun{sussman1994level}.

Finally, as usual, a mirror boundary condition is imposed at $r=0$. For the remaining three boundaries, we also adopt mirror boundary conditions. This latter choice does not affect the capillary instability as long as the computational boundary in the $r$~direction is well-separated from the interface between the two fluids. Note also that there are large thermal and viscosity gradients in the $z$~direction. Therefore, a mirror boundary condition is justified at the low-temperature end where the interface motion is very slow. At the high-temperature boundary, the fluid column has already broken into several droplets far upstream, so the mirror boundary condition does not affect the breakup process either.

\bibliography{capillary}

\end{document}